\documentclass{article}
\usepackage{amssymb}
\usepackage{amsfonts}
\usepackage{amsmath}
\usepackage[doublespacing]{setspace}

\input{tcilatex}

\begin{document}

\title{Radial power-law position-dependent mass; Cylindrical coordinates,
separability, and spectral signatures}
\author{Omar Mustafa \\
Department of Physics, Eastern Mediterranean University, \\
G Magusa, North Cyprus, Mersin 10,Turkey\\
E-mail: omar.mustafa@emu.edu.tr, \\
\ Tel: +90 392 630 1314,\\
\ Fax: +90 392 3651604}
\maketitle

\begin{abstract}
We discuss the separability of the position-dependent mass Hamiltonian in
cylindrical coordinates in the framework of a radial power-law
position-dependent mass. We consider two particular radial mass settings; a
harmonic oscillator type, and a Coulombic type. We subject the radial
harmonic oscillator type mass to a radial harmonic oscillator potential and
the radial Coulombic mass to a radial Coulombic potential. Azimuthal
symmetry is assumed and spectral signatures of various $z$-dependent
interaction potentials are reported.

\medskip PACS codes: 03.65.Ge, 03.65.Ca

Keywords: Power-law, Position-dependent-mass, cylindrical coordinates,
separability, exact solvability, spectral signatures
\end{abstract}

\section{Introduction}

The von Roos Hamiltonian [1] is known to describe quantum particles with
position-dependent-mass (PDM), $M\left( \vec{r}\right) =m_{\circ }m\left( 
\vec{r}\right) $. Over the last few decades, the position-dependent-mass
Hamiltonians have inspired research attention [2-32] because of their
applicability in the study of many-body problem, semiconductors, quantum
dots, quantum liquids, etc. The kinetic energy operator in the von Roos
Hamiltonian (with $m_{\circ }=\hbar =1$ units) 
\begin{equation}
H=-\frac{1}{4}\left[ m\left( \vec{r}\right) ^{\gamma }\vec{\nabla}m\left( 
\vec{r}\right) ^{\beta }\mathbf{\cdot }\vec{\nabla}m\left( \vec{r}\right)
^{\alpha }+m\left( \vec{r}\right) ^{\alpha }\;\vec{\nabla}m\left( \vec{r}%
\right) ^{\beta }\mathbf{\cdot }\vec{\nabla}m\left( \vec{r}\right) ^{\gamma }%
\right] +V\left( \vec{r}\right) ,
\end{equation}%
admits an ordering ambiguity manifested by the non-uniqueness representation
of the kinetic energy operator. Which would, in effect, introduce a profile
change in the effective potential as the values of the parameters $\alpha $, 
$\beta $, and $\gamma $ change [cf., e.g., 25-29]. Here, $\alpha $, $\beta $%
, and $\gamma $ are called the von Roos ordering ambiguity parameters
satisfying the von Roos constraint $\alpha +\beta +\gamma =-1$ .
Nevertheless, an interesting and comprehensive background on the issue of
consistency and usefulness of the position-dependent mass Schr\"{o}dinger
equation was given by L\.{e}vy-Leblond [32]. Therein, his work is devoted to
sustaining and strengthening the conclusions that not only the use of
position-dependent mass gives correct approximation, but its also a
conceptually consistent approach.

It is however constructive to mention that the continuity conditions at the
abrupt heterojunction between two crystals implied that $\alpha =\gamma $,
otherwise for $\alpha \neq \gamma $ the wavefunctions vanish at the
boundaries and the heterojunction plays the role of impenetrable barrier
(cf., e.g., Mustafa and Mazharimousavi [10] and Koc et al. in [28]).
Eliminating in the process, the Gora's and Williams' ($\beta =\gamma =0,$ $%
\alpha =-1$), and Li's and Kuhn's ($\beta =\gamma =-1/2,$ $\alpha =0$) known
parametric sets. \ Moreover, Dutra's and Almeida's [9] reliability test
classifies the parametric set of Ben Daniel and Duke ($\alpha =\gamma =0,$ $%
\beta =-1$) as a set to-be-discarded for it yields imaginary eigenvalues.
This would leave us with Zhu's and Kroemer's ($\alpha =\gamma =-1/2,$ $\beta
=0$) and Mustafa's and Mazharimousavi's ($\alpha =\gamma =-1/4,$ $\beta
=-1/2 $) ordering ambiguity parameters that are classified as "good"
parametric sets, so to speak. Nevertheless, we have observed (cf., e. g.,
[29]) that the physical and/or mathematical admissibility of a given
ambiguity parametric set depends also on the form of the
position-dependent-mass and/or the form of the interaction potential. In the
forthcoming methodical proposal, we shall work with the ambiguity parameters
as they are without any classification as to which set is "good" or
"to-be-discarded".

Very recently, Mustafa [31] has considered the von Roos Hamiltonian (1)
using cylindrical coordinates. Therein, we sought some manifestly feasible
separability through the suggestion that the position-dependent-mass is only
radial-dependent (i.e., $m\left( \vec{r}\right) =m_{\circ }M\left( \rho
,\varphi ,z\right) =M\left( \rho ,\varphi ,z\right) =M\left( \rho \right)
=1/\rho ^{2}$), where azimuthal symmetrization is granted through a proper
assumption of the interaction potential. The spectral signatures of
different $z$-dependent interaction potential settings on the radial
Coulombic and radial harmonic oscillator interaction potentials' spectra are
reported for impenetrable walls at $z=0$ and $z=L$, for a Morse, for a
non-Hermitian $\mathcal{PT}$-symmetrized Scarf II, and for a non-Hermitian $%
\mathcal{PT}$-symmetrized Samsonov interaction models.

In this work, we offer a parallel azimuthal symmetrization along with a more
general (though still only radial-dependent) power-law-type
position-dependent-mass (i.e., $M\left( \rho ,\varphi ,z\right) =M\left(
\rho \right) =b\rho ^{2\upsilon +1}/2$). Obviously, a $\upsilon =-3/2$ and $%
b=2$ yield $M\left( \rho \right) \sim 1/\rho ^{2}$ which is, under the
current forthcoming settings, a special case of $M\left( \rho \right) =b\rho
^{2\upsilon +1}/2$ that shall not be repeated here. Instead, we shall use $%
\upsilon =-1$ and $\upsilon =1/2$ that yield quantum particles endowed with
position-dependent masses of a Coulombic-type, $M\left( \rho \right) =b\rho
^{-1}/2$, and a harmonic oscillator type, $M\left( \rho \right) =b\rho
^{2}/2 $, respectively. To the best of our knowledge, such
position-dependent mass settings have not been considered elsewhere.

To make this work self-contained, we recollect (in section 2) the most
relevant and vital relations (namely, equations (2)-(5) below) that have
been readily reported in [31] for cylindrical coordinates separability and
exact solvability of the PDM-Hamiltonian (1). In the same section, we
discuss the separability in the framework of a manifestly radial power-law
position-dependent mass and contemplate on the feasible separabilities. In
section 3, we consider two particular radial mass settings; a harmonic
oscillator type, $M\left( \rho ,\varphi ,z\right) =M\left( \rho \right)
=g\left( \rho \right) =b\rho ^{2}/2$, and a Coulombic type, $M\left( \rho
,\varphi ,z\right) =M\left( \rho \right) =g\left( \rho \right) =b\rho
^{-1}/2 $. We subject the radial harmonic oscillator type mass to move under
the influence of a radial harmonic oscillator potential field $\tilde{V}%
\left( \rho \right) =a^{2}\rho ^{2}/4$ and the radial Coulombic mass to a
radial Coulombic potential $\tilde{V}\left( \rho \right) =-2\,\tilde{A}/\rho 
$. The spectral signatures of (i) two impenetrable walls at $z=0$ and $z=L$
provided by the potential well $\tilde{V}\left( z\right) =0$ for $0<z<L$ and 
$\tilde{V}\left( z\right) =\infty $ elsewhere, (ii) a Morse type [31]
interaction $\tilde{V}\left( z\right) =D\ \left( e^{-2\epsilon
z}-2e^{-\epsilon z}\right) ;$ $D>0$, and (iii) a trigonometric Rosen-Morse
[32] potential $\tilde{V}\left( z\right) =U_{\circ }\cot ^{2}\left( \pi
z/d\right) ;z\in \left[ 0,d\right] $, are reported in the same section. Our
concluding remarks are given in section 4.

\section{Cylindrical coordinates and radial power-law PDM framework}

Following our recent work [31] on cylindrical coordinates separability and
exact solvability of the PDM-Hamiltonian (1), we again consider the
position-dependent-mass and the interaction potential to take the forms $%
m\left( \vec{r}\right) \equiv $ $M\left( \rho ,\varphi ,z\right) =g\left(
\rho \right) f\left( \varphi \right) k\left( z\right) $ and $V\left( \vec{r}%
\right) \equiv V\left( \rho ,\varphi ,z\right) $, respectively. We have
shown (see Mustafa [31] for more details on this issue) that the
corresponding PDM-Schr\"{o}dinger equation $\left[ H-E\right] \Psi \left(
\rho ,\varphi ,z\right) =0$ with%
\begin{equation}
\Psi \left( \rho ,\varphi ,z\right) =R\left( \rho \right) \Phi \left(
\varphi \right) Z\left( z\right) ;\text{ }\rho \in \left( 0,\infty \right)
,\ \varphi \in \left( 0,2\pi \right) ,z\in \left( -\infty ,\infty \right) ,
\end{equation}%
would imply%
\begin{eqnarray}
0 &=&2g\left( \rho \right) f\left( \varphi \right) k\left( z\right) \left[
E-V\left( \rho ,\varphi ,z\right) \right]  \notag \\
&&+\left[ \frac{R^{\prime \prime }\left( \rho \right) }{R\left( \rho \right) 
}-\left( \frac{g^{\prime }\left( \rho \right) }{g\left( \rho \right) }-\frac{%
1}{\rho }\right) \frac{R^{\prime }\left( \rho \right) }{R\left( \rho \right) 
}\right.  \notag \\
&&+\left. \frac{\zeta }{2}\left( \frac{g^{\prime }\left( \rho \right) }{%
g\left( \rho \right) }\right) ^{2}-\frac{\left( \beta +1\right) }{2}\left( 
\frac{g^{\prime }\left( \rho \right) }{\rho g\left( \rho \right) }+\frac{%
g^{^{\prime \prime }}\left( \rho \right) }{g\left( \rho \right) }\right) %
\right]  \notag \\
&&+\left[ \frac{Z^{\prime \prime }\left( z\right) }{Z\left( z\right) }-\frac{%
k^{\prime }\left( z\right) }{k\left( z\right) }\frac{Z^{\prime }\left(
z\right) }{Z\left( z\right) }+\frac{\zeta }{2}\left( \frac{k^{\prime }\left(
z\right) }{k\left( z\right) }\right) ^{2}-\frac{\left( \beta +1\right) }{2}%
\frac{k^{^{\prime \prime }}\left( z\right) }{k\left( z\right) }\right] 
\notag \\
&&+\frac{1}{\rho ^{2}}\left[ \frac{\Phi ^{\prime \prime }\left( \varphi
\right) }{\Phi \left( \varphi \right) }-\frac{f^{\prime }\left( \varphi
\right) }{f\left( \varphi \right) }\frac{\Phi ^{\prime }\left( \varphi
\right) }{\Phi \left( \varphi \right) }+\frac{\zeta }{2}\left( \frac{%
f^{\prime }\left( \varphi \right) }{f\left( \varphi \right) }\right) ^{2}-%
\frac{\left( \beta +1\right) }{2}\frac{f^{^{\prime \prime }}\left( \varphi
\right) }{f\left( \varphi \right) }\right]
\end{eqnarray}%
where%
\begin{equation}
\zeta =\alpha \left( \alpha -1\right) +\gamma \left( \gamma -1\right) -\beta
\left( \beta +1\right) .
\end{equation}%
To facilitate and ease separability, we have suggested that the interaction
potential satisfies an obviously "manifested-by-equation (3)" general
identity of the form%
\begin{equation}
2MV\left( \rho ,\varphi ,z\right) =2g\left( \rho \right) f\left( \varphi
\right) k\left( z\right) V\left( \rho ,\varphi ,z\right) =\tilde{V}\left(
\rho \right) +\tilde{V}\left( z\right) +\frac{1}{\rho ^{2}}\tilde{V}\left(
\varphi \right) .
\end{equation}%
Hereby, we may remind the reader that in [31\} we have used $g\left( \rho
\right) =1/\rho ^{2}$ along with $f\left( \varphi \right) =1=k\left(
z\right) $ as one of the options that secured separability of the problem at
hand.

In the search for a more general recipe, however, we choose to eliminate the
first-order derivatives $Z^{\prime }\left( z\right) $, $\Phi ^{\prime
}\left( \varphi \right) $, and $R^{\prime }\left( \rho \right) $. At this
point, the elimination of the first-order derivatives of $Z\left( z\right) $
and $\Phi \left( \varphi \right) $ is achieved through the substitutions 
\begin{equation}
Z\left( z\right) =\sqrt{k\left( z\right) }\tilde{Z}\left( z\right) \text{
and }\Phi \left( \varphi \right) =\sqrt{f\left( \varphi \right) }\tilde{\Phi}%
\left( \varphi \right) ,
\end{equation}%
to imply that%
\begin{equation}
\frac{Z^{\prime \prime }\left( z\right) }{Z\left( z\right) }-\frac{k^{\prime
}\left( z\right) }{k\left( z\right) }\frac{Z^{\prime }\left( z\right) }{%
Z\left( z\right) }=-\frac{3}{4}\left( \frac{k^{\prime }\left( z\right) }{%
k\left( z\right) }\right) ^{2}+\frac{1}{2}\frac{k^{^{\prime \prime }}\left(
z\right) }{k\left( z\right) }+\frac{\tilde{Z}^{\prime \prime }\left(
z\right) }{\tilde{Z}\left( z\right) }
\end{equation}%
and%
\begin{equation}
\frac{\Phi ^{\prime \prime }\left( \varphi \right) }{\Phi \left( \varphi
\right) }-\frac{f^{\prime }\left( \varphi \right) }{f\left( \varphi \right) }%
\frac{\Phi ^{\prime }\left( \varphi \right) }{\Phi \left( \varphi \right) }=-%
\frac{3}{4}\left( \frac{f^{\prime }\left( \varphi \right) }{f\left( \varphi
\right) }\right) ^{2}+\frac{1}{2}\frac{f^{^{\prime \prime }}\left( \varphi
\right) }{f\left( \varphi \right) }+\frac{\tilde{\Phi}^{\prime \prime
}\left( \varphi \right) }{\tilde{\Phi}\left( \varphi \right) }
\end{equation}%
Whereas, the elimination of the first-order derivative of $R\left( \rho
\right) $ may be sought through the substitutions%
\begin{equation}
R\left( \rho \right) =\rho ^{\upsilon }U\left( \rho \right) \text{ and }%
g\left( \rho \right) =\frac{b}{2}\rho ^{2\upsilon +1};\upsilon ,b\in 
\mathbb{R}
,
\end{equation}%
(with the restriction that $b$ is a non-zero constant to avoid triviality)
to imply that%
\begin{equation}
\frac{R^{\prime \prime }\left( \rho \right) }{R\left( \rho \right) }-\left( 
\frac{g^{\prime }\left( \rho \right) }{g\left( \rho \right) }-\frac{1}{\rho }%
\right) \frac{R^{\prime }\left( \rho \right) }{R\left( \rho \right) }=\frac{%
U^{\prime \prime }\left( \rho \right) }{U\left( \rho \right) }-\frac{%
\upsilon \left( \upsilon +1\right) }{\rho ^{2}}.
\end{equation}%
It should be noted here that the choice of $g\left( \rho \right) =b\rho
^{2\upsilon +1}/2$ in (9) is manifestly mandated by the elimination of the
first-order derivative of $U\left( \rho \right) $.

Under such settings, equation (3) would read%
\begin{eqnarray}
0 &=&b\rho ^{2\upsilon +1}f\left( \varphi \right) k\left( z\right) E  \notag
\\
&&+\left[ \frac{U^{\prime \prime }\left( \rho \right) }{U\left( \rho \right) 
}+\frac{\left( 2\upsilon +1\right) ^{2}\left[ \zeta -\beta -1\right]
-2\upsilon \left( \upsilon +1\right) }{2\rho ^{2}}-\tilde{V}\left( \rho
\right) \right]  \notag \\
&&+\left[ \frac{\tilde{Z}^{\prime \prime }\left( z\right) }{\tilde{Z}\left(
z\right) }+\frac{\left( 2\zeta -3\right) }{4}\left( \frac{k^{\prime }\left(
z\right) }{k\left( z\right) }\right) ^{2}-\frac{\beta }{2}\frac{k^{^{\prime
\prime }}\left( z\right) }{k\left( z\right) }-\tilde{V}\left( z\right) %
\right]  \notag \\
&&+\frac{1}{\rho ^{2}}\left[ \frac{\tilde{\Phi}^{\prime \prime }\left(
\varphi \right) }{\tilde{\Phi}\left( \varphi \right) }+\frac{\left( 2\zeta
-3\right) }{4}\left( \frac{f^{\prime }\left( \varphi \right) }{f\left(
\varphi \right) }\right) ^{2}-\frac{\beta }{2}\frac{f^{^{\prime \prime
}}\left( \varphi \right) }{f\left( \varphi \right) }-\tilde{V}\left( \varphi
\right) \right] .
\end{eqnarray}%
Hereby, it should mentioned that the case where $\upsilon =-3/2$, $b=2$
(i.e., $g\left( \rho \right) =1/\rho ^{2}$) and $k\left( z\right) =1=f\left(
\varphi \right) $ is the case we have considered in [31]. It is now just a
special case of the current, though more general, recipe $g\left( \rho
\right) =b\rho ^{2\upsilon +1}/2;\upsilon ,b\in 
\mathbb{R}
$. The results and examples reported therein are recoverable and hold true
for the current work, therefore.

Yet, an obvious manifestation of the energy term $b\rho ^{2\upsilon
+1}f\left( \varphi \right) k\left( z\right) E$ towards separability of (11)
is that, in addition to the three feasible separable cases $f\left( \varphi
\right) =1=k\left( z\right) $, $k\left( z\right) =1=g\left( \rho \right) $,
and $f\left( \varphi \right) =1=g\left( \rho \right) $ (reported in [31]),
one finds two more feasibly separable cases. They are, for $\upsilon =-3/2$, 
$k\left( z\right) =1$, $f\left( \varphi \right) \neq 1$ (which, in turn,
would break azimuthal symmetry) and for $\upsilon =-3/2$, $k\left( z\right)
\neq 1$, $f\left( \varphi \right) =1$. Therefore, the separability of (11)
may be facilitated by the forms of the position-dependent mass and of the
interaction potential $V\left( \rho ,\varphi ,z\right) $.

To secure azimuthal symmetrization of the problem at hand we substitute $%
\tilde{V}\left( \varphi \right) =0$ and $f\left( \varphi \right) =1$.
Moreover, we choose $k\left( z\right) =1$ to imply that%
\begin{equation}
\frac{\tilde{\Phi}^{\prime \prime }\left( \varphi \right) }{\tilde{\Phi}%
\left( \varphi \right) }=k_{\varphi }^{2}\text{ ; }k_{\varphi }^{2}=-m^{2}%
\text{ ; }\left\vert m\right\vert =0,1,2,\cdots ,
\end{equation}%
\begin{equation}
\left[ -\partial _{z}^{2}+\tilde{V}\left( z\right) \right] \,\tilde{Z}\left(
z\right) =k_{z}^{2}\,\tilde{Z}\left( z\right) ,
\end{equation}%
and%
\begin{equation}
\left[ -\partial _{\rho }^{2}+\frac{\tilde{\ell}_{\upsilon }^{2}-1/4}{\rho
^{2}}+\tilde{V}\left( \rho \right) -b\rho ^{2\upsilon +1}E\right] U\left(
\rho \right) =-k_{z}^{2}U\left( \rho \right) .
\end{equation}%
Where $m$ is the magnetic quantum number and%
\begin{equation}
\left\vert \tilde{\ell}_{\upsilon }\right\vert =\sqrt{\upsilon \left(
\upsilon +1\right) +m^{2}+\frac{1}{4}-\frac{\left( 2\upsilon +1\right) ^{2}%
\left[ \zeta -\beta -1\right] }{2}}.
\end{equation}%
is an irrational magnetic quantum number. Hereby, it obvious that the
substitutions of $\upsilon =-3/2$ and $b=2$ in (14) would inspire a re-scale
of the form%
\begin{equation}
\tilde{\ell}_{-3/2}^{2}=\ell ^{2}+2E=\left( m^{2}+3\right) -2\left( \zeta
-\beta \right) +2E,
\end{equation}%
so that our results in [31] for the radial Coulombic $\tilde{V}\left( \rho
\right) =-2/\rho $ and the radial harmonic oscillator $\tilde{V}\left( \rho
\right) =a^{2}\rho ^{2}/4$ (equations (23) and (24) in [31], respectively)
are safely reproduced. Therefore, the corresponding spectral signatures of $%
\tilde{V}\left( z\right) $ interaction potentials for $\upsilon =-3/2$ and $%
b=2$ (i.e, $g\left( \rho \right) =\rho ^{-2}$) are reported therein [31] and
shall not be repeated here.

\section{Two particular radial settings; $g\left( \protect\rho \right) =b%
\protect\rho ^{2}/2$ and $g\left( \protect\rho \right) =b\protect\rho %
^{-1}/2 $}

In this section. we consider the position-dependent mass $M\left( \rho
,\varphi ,z\right) =g\left( \rho \right) $ to indulge a radial harmonic
oscillator $g\left( \rho \right) =b\rho ^{2}/2$ (i.e., $\upsilon =1/2$) and
the radial Coulombic $g\left( \rho \right) =b\rho ^{-1}/2$ (i.e., $\upsilon
=-1$) forms. For the sake of keeping this work simple and instructive, we
shall consider the radial harmonic oscillator $g\left( \rho \right) =b\rho
^{2}/2$ accompanied by a radial harmonic oscillator type interaction $\tilde{%
V}\left( \rho \right) =a^{2}\rho ^{2}/4$ and the radial Coulombic $g\left(
\rho \right) =b\rho ^{-1}/2$ accompanied by a radial Coulombic $\tilde{V}%
\left( \rho \right) =-2\,\tilde{A}/\rho $. We shall moreover report the
spectral signatures of different $\tilde{V}\left( z\right) $ potentials on
the overall spectrum.

\subsection{The radial harmonic oscillator $g\left( \protect\rho \right) =b%
\protect\rho ^{2}/2$}

The choice of $\upsilon =1/2$ along with $\tilde{V}\left( \rho \right)
=a^{2}\rho ^{2}/4$ would imply that equation (14) reads%
\begin{equation}
\left[ -\partial _{\rho }^{2}+\frac{\tilde{\ell}_{1/2}^{2}-1/4}{\rho ^{2}}+%
\frac{\left( a^{2}-4bE\right) }{4}\rho ^{2}\right] U\left( \rho \right)
=-k_{z}^{2}U\left( \rho \right) ,
\end{equation}%
and (15), in turn, yields%
\begin{equation}
\left\vert \tilde{\ell}_{1/2}\right\vert =\sqrt{\left( m^{2}+3\right)
-2\left( \zeta -\beta \right) }.
\end{equation}%
Obviously, Eq.(17) has exact eigenvalues in the form%
\begin{equation}
k_{z}^{2}=-\sqrt{\left( a^{2}-4bE\right) }\left[ 2n_{\rho }+\left\vert 
\tilde{\ell}_{1/2}\right\vert +1\right] ^{2},
\end{equation}%
and implies that%
\begin{equation}
E=\frac{a^{2}}{4b}-\frac{1}{4b}\left[ \frac{k_{z}^{2}}{2n_{\rho }+\sqrt{%
\left( m^{2}+3\right) -2\left( \zeta -\beta \right) }+1}\right] ^{2}.
\end{equation}%
We observe that an auxiliary constraint%
\begin{equation}
\left( \zeta -\beta \right) =\alpha \left( \alpha -1\right) +\gamma \left(
\gamma -1\right) -\beta \left( \beta +2\right) \leq \left( m^{2}+3\right) /2
\end{equation}%
on the ambiguity parameters is manifested here by the requirement that $E\in 
\mathbb{R}
$.

\subsubsection{Spectral signatures of some $\tilde{V}\left( z\right) $
potentials on the radial harmonic oscillator spectrum}

Recollect [31] that if our PDM-particle is trapped to move between two
impenetrable walls at $z=0$ and $z=L$ under the influence of a%
\begin{equation}
\tilde{V}\left( z\right) =\left\{ 
\begin{tabular}{ll}
$0$ & ; $0<z<L$ \\ 
$\infty $ & ; elsewhere%
\end{tabular}%
\right. ,
\end{equation}%
one would find that $K_{z}=n_{z}\pi /L\ ,$ $n_{z}=1,2,3,\cdots $ (see [31]
for more details on this issue). This would, in effect, give the spectral
signature of $\tilde{V}\left( z\right) $ of (22) on the overall spectrum%
\begin{equation}
E=\frac{a^{2}}{4b}-\frac{1}{4b}\left[ \frac{\left( n_{z}\pi /L\right) ^{2}}{%
2n_{\rho }+\sqrt{\left( m^{2}+3\right) -2\left( \zeta -\beta \right) }+1}%
\right] ^{2}
\end{equation}%
for a PDM particle of $M\left( \rho ,\varphi ,z\right) =M\left( \rho \right)
=b\rho ^{2}/2$ moving in a potential of the form%
\begin{equation}
V\left( \rho ,\varphi ,z\right) =\frac{a^{2}}{4b}+\left\{ 
\begin{tabular}{ll}
$0$ & ; $0<z<L$ \\ 
$\infty $ & ; elsewhere%
\end{tabular}%
\right. .
\end{equation}

Next, let us subject this PDM particle to move in a Morse type [31]
interaction $\tilde{V}\left( z\right) =D\ \left( e^{-2\epsilon
z}-2e^{-\epsilon z}\right) ;$ $D>0$. In this case%
\begin{equation}
k_{z}^{2}=\left( \frac{\sqrt{D}}{\epsilon }-\tilde{n}_{z}-\frac{1}{2}\right)
,\text{ }\tilde{n}_{z}=0,1,2,3,\cdots
\end{equation}%
Therefore, a PDM quantum particle endowed with $M\left( \rho ,\varphi
,z\right) =M\left( \rho \right) =b\rho ^{2}/2$ and subjected to an
interaction potential of the form%
\begin{equation}
V\left( \rho ,\varphi ,z\right) =\frac{a^{2}}{4b}+\frac{D}{b\rho ^{2}}\
\left( e^{-2\epsilon z}-2e^{-\epsilon z}\right) ;D>0
\end{equation}%
would admit exact energy eigenvalues given by%
\begin{equation}
E=\frac{a^{2}}{4b}-\frac{1}{4b}\left[ \frac{\left( \sqrt{D}/\epsilon -\tilde{%
n}_{z}-\frac{1}{2}\right) }{2n_{\rho }+\sqrt{\left( m^{2}+3\right) -2\left(
\zeta -\beta \right) }+1}\right] ^{2}.
\end{equation}

Now, let $M\left( \rho ,\varphi ,z\right) =M\left( \rho \right) =b\rho
^{2}/2 $ move under the influence of a trigonometric Rosen-Morse potential $%
\tilde{V}\left( z\right) =U_{\circ }\cot ^{2}\left( \pi z/d\right) ;z\in %
\left[ 0,d\right] $. Where $U_{\circ }$ and $d$ are two positive parameters.
In this case, 
\begin{equation}
V\left( \rho ,\varphi ,z\right) =\frac{a^{2}}{4b}+\frac{U_{\circ }}{b\rho
^{2}}\ \cot ^{2}\left( \pi z/d\right) \,;\,z\in \left[ 0,d\right] ,
\end{equation}%
\begin{equation}
k_{z}^{2}=\frac{1}{d^{2}}\left[ Cd+\tilde{n}_{z}\pi \right] ^{2}-U_{\circ }%
\text{ ; }C=\frac{\pi }{2d}\left( 1+\sqrt{1+\frac{4U_{\circ }d^{2}}{\pi ^{2}}%
}\right) ,
\end{equation}%
(see Ma et al [32] for more details, notice that one should consider $2\mu
=\hbar =1$ of Ma as proper parametric mapping into our settings) and%
\begin{equation}
E=\frac{a^{2}}{4b}-\frac{1}{4b}\left[ \frac{\left[ Cd+\tilde{n}_{z}\pi %
\right] ^{2}/d^{2}-U_{\circ }}{2n_{\rho }+\sqrt{\left( m^{2}+3\right)
-2\left( \zeta -\beta \right) }+1}\right] ^{2}.
\end{equation}

\subsection{The Radial Coulombic $g\left( \protect\rho \right) =b\protect%
\rho ^{-1}/2$}

Now consider the PDM-particle to have a radial Coulombic-type mass of the
form $M\left( \rho ,\varphi ,z\right) =M\left( \rho \right) =b\rho ^{-1}/2$,
(i.e., $\upsilon =-1$) and subjected to move in a radial Coulombic potential 
$\tilde{V}\left( \rho \right) =-2\,\tilde{A}/\rho $. In this case,%
\begin{equation}
V\left( \rho ,\varphi ,z\right) =-\frac{\tilde{A}}{b}+\frac{\rho }{b}\tilde{V%
}\left( z\right) ,
\end{equation}%
and equation (14) yields%
\begin{equation}
\left[ -\partial _{\rho }^{2}+\frac{\tilde{\ell}_{-1}^{2}-1/4}{\rho ^{2}}-%
\frac{2\left( \tilde{b}E+\tilde{A}\right) }{\rho }^{2}\right] U\left( \rho
\right) =-k_{z}^{2}U\left( \rho \right) ;\,\tilde{b}=b/2,\,
\end{equation}%
with%
\begin{equation}
\left\vert \tilde{\ell}_{-1}\right\vert =\sqrt{\left( m^{2}+3/4\right)
-\left( \zeta -\beta \right) /2},
\end{equation}%
and%
\begin{equation}
k_{z}=\pm \frac{\tilde{b}E+\tilde{A}}{\left( n_{\rho }+\left\vert \tilde{\ell%
}_{-1}\right\vert +1\right) }.
\end{equation}%
Which, in turn, results 
\begin{equation}
E=\pm \frac{k_{z}}{\tilde{b}}\left( n_{\rho }+\sqrt{\left( m^{2}+3/4\right)
-\left( \zeta -\beta \right) /2}+1\right) -\frac{\tilde{A}}{\tilde{b}},
\end{equation}%
with the auxiliary constraint%
\begin{equation}
\left( \zeta -\beta \right) =\alpha \left( \alpha -1\right) +\gamma \left(
\gamma -1\right) -\beta \left( \beta +2\right) \leq \left( 2m^{2}+3/2\right)
,
\end{equation}%
on the ambiguity parameters that secures the reality of $E$. Nevertheless,
two branches of energies are obviously obtained. Moreover, the spectral
signature of $k_{z}$ on the overall spectrum is obtained through the
solution of equation (13).

\subsubsection{Spectral signatures of some $\tilde{V}\left( z\right) $
potentials on the radial Coulombic spectrum}

If we subject our radial Coulombic PDM-particle $M\left( \rho ,\varphi
,z\right) =M\left( \rho \right) =b\rho ^{-1}/2$ to move in $V\left( \rho
,\varphi ,z\right) =-\tilde{A}/b+\rho \tilde{V}\left( z\right) /b,$where $%
\tilde{V}\left( z\right) $ is given by (22), it will admit exact energies of
the form%
\begin{equation}
E=\pm \frac{n_{z}\pi }{\tilde{b}L}\left( n_{\rho }+\sqrt{\left(
m^{2}+3/4\right) -\left( \zeta -\beta \right) /2}+1\right) -\frac{\tilde{A}}{%
\tilde{b}};\,n_{z}=1,2,3,\cdots .
\end{equation}

Moreover, if this PDM-particle is subjected to move in a Morse type [31]
interaction $\tilde{V}\left( z\right) =D\ \left( e^{-2\epsilon
z}-2e^{-\epsilon z}\right) ;$ $D>0$. In this case,%
\begin{equation}
V\left( \rho ,\varphi ,z\right) =-\frac{\tilde{A}}{b}+\frac{\rho }{b}D\
\left( e^{-2\epsilon z}-2e^{-\epsilon z}\right) ;D>0,
\end{equation}%
and the exact energies are of the form%
\begin{equation}
E=\pm \frac{1}{\tilde{b}}\sqrt{\left( \frac{\sqrt{D}}{\epsilon }-\tilde{n}%
_{z}-\frac{1}{2}\right) }\left( n_{\rho }+\sqrt{\left( m^{2}+3/4\right)
-\left( \zeta -\beta \right) /2}+1\right) -\frac{\tilde{A}}{\tilde{b}},
\end{equation}%
Where$\,\tilde{n}_{z}=0,1,2,3,\cdots $. Obviously, the condition $\left( 
\sqrt{D}/\epsilon -\tilde{n}_{z}-\frac{1}{2}\right) >0$ is manifested here
and ought to be enforced, otherwise complex pairs of energy eigenvalues are
obtained in the process.

Next, let $M\left( \rho ,\varphi ,z\right) =M\left( \rho \right) =b\rho
^{-1}/2$ move under the influence of a trigonometric Rosen-Morse potential $%
\tilde{V}\left( z\right) =U_{\circ }\cot ^{2}\left( \pi z/d\right) ;z\in %
\left[ 0,d\right] $. Then,%
\begin{equation}
V\left( \rho ,\varphi ,z\right) =-\frac{\tilde{A}}{b}+\frac{\rho }{b}%
U_{\circ }\cot ^{2}\left( \pi z/d\right) ;z\in \left[ 0,d\right] ,
\end{equation}%
and%
\begin{equation}
E=\pm \frac{\sqrt{\left[ Cd+\tilde{n}_{z}\pi \right] ^{2}-U_{\circ }d^{2}}}{%
\tilde{b}d}\left( n_{\rho }+\sqrt{\left( m^{2}+3/4\right) -\left( \zeta
-\beta \right) /2}+1\right) -\frac{\tilde{A}}{\tilde{b}}.
\end{equation}

\section{Concluding remarks}

We have recollected the most relevant and vital relations (equations (2)-(5)
above) that have been readily reported by Mustafa [31] for cylindrical
coordinates separability of the PDM-Hamiltonian in (1), where the
PDM-setting was considered in the form $M\left( \rho ,\varphi ,z\right)
=g\left( \rho \right) f\left( \varphi \right) k\left( z\right) =g\left( \rho
\right) =1/\rho ^{2}$ under azimuthally symmetric settings.

In this work, however, we offered a more general power-law radial
position-dependent mass recipe $M\left( \rho ,\varphi ,z\right) =g\left(
\rho \right) =b\rho ^{2\upsilon +1}/2;\upsilon ,b\in 
\mathbb{R}
$, within which $M\left( \rho ,\varphi ,z\right) =g\left( \rho \right)
=1/\rho ^{2}$ of [31] represents a special case (the results and examples
reported therein hold true and yet document additional examples on the
applicability of the current methodical proposal, therefore). Moreover, the
structure of the position-dependent energy term $b\rho ^{2\upsilon
+1}f\left( \varphi \right) k\left( z\right) E$ in (11) suggests that there
are five feasible cases towards separability; (i) $f\left( \varphi \right)
=1=k\left( z\right) $, (ii) $k\left( z\right) =1=g\left( \rho \right) $,
(iii) $f\left( \varphi \right) =1=g\left( \rho \right) $, (iv) $\upsilon
=-3/2$, $k\left( z\right) \neq 1$, $f\left( \varphi \right) =1$, and (v) $%
\upsilon =-3/2$, $k\left( z\right) =1$, $f\left( \varphi \right) \neq 1$
(which would break azimuthal symmetry, of course). Therefore, the
separability of (3) may be facilitated by the forms of the
position-dependent mass and the interaction potential $V\left( \rho ,\varphi
,z\right) $. These are not the only cases to secure separability of (3), so
to speak.

We have considered two particular mass settings; a radial harmonic
oscillator type, $M\left( \rho ,\varphi ,z\right) =M\left( \rho \right)
=g\left( \rho \right) =b\rho ^{2}/2$, and a radial Coulombic type, $M\left(
\rho ,\varphi ,z\right) =M\left( \rho \right) =g\left( \rho \right) =b\rho
^{-1}/2$. We have observed that for the Coulombic case two branches of
energies are obtained, each of which is a "mirror-reflection" of the other
about the zero-energy axis. Moreover, when we subjected the radial harmonic
oscillator mass to a radial harmonic oscillator potential $\tilde{V}\left(
\rho \right) =a^{2}\rho ^{2}/4$ and the radial Coulombic mass to a radial
Coulombic potential $\tilde{V}\left( \rho \right) =-2\,\tilde{A}/\rho $,
only constant shifts in the energies where observed (i.e., a shift $\left(
a^{2}/4b\right) $ for the radial harmonic oscillator mass and $\left( -%
\tilde{A}/\tilde{b}\right) $ for the radial Coulombic mass, documented in
(20) and (35), respectively). That is, the radial interaction potentials $%
\tilde{V}\left( \rho \right) $ considered for the two over simplified
examples here provided no quantization recipe at all (i.e., they have only
introduced constant shifts to the energies but not discrete quantum energy
shifts). This is because the form of the general interaction potential $%
V\left( \rho ,\varphi ,z\right) $ we have adopted in (5).

Yet, auxiliary constraints on the ambiguity parameters (see (21) for the
harmonic oscillator and (36) for the Coulombic) are observed mandatory to
secure the reality of $E$. Hereby, if \ $m=0$ is considered in (21) and (36)
as a reference test, then one would observe that only the Gora's and
Williams' ambiguity parametric set ($\beta =\gamma =0,$ $\alpha =-1$) fails
to provide real energies (i.e., $\sqrt{3/2-\left( \zeta -\beta \right) }\in 
\mathbb{C}
$) . We contemplate that more auxiliary constraints on the ambiguity
parameters should be anticipated for different, though exactly solvable,
power-law type radial masses (within our methodical proposal, of course).
Furthermore, the spectral signatures of different $\tilde{V}\left( z\right) $
interactions on the overall spectrum are also reported. Namely, the spectral
signatures of (i) two impenetrable walls at $z=0$ and $z=L$ provided by the
potential well $\tilde{V}\left( z\right) =0$ for $0<z<L$ and $\tilde{V}%
\left( z\right) =\infty $ elsewhere, (ii) a Morse type [31] interaction $%
\tilde{V}\left( z\right) =D\ \left( e^{-2\epsilon z}-2e^{-\epsilon z}\right)
;$ $D>0$, and (iii) a trigonometric Rosen-Morse [32] potential $\tilde{V}%
\left( z\right) =U_{\circ }\cot ^{2}\left( \pi z/d\right) ;z\in \left[ 0,d%
\right] $.

\end{document}